\newcommand\bm[1]{\mbox{\boldmath$#1$}}
\def\sqr#1#2{{\vcenter{\vbox{\hrule height.#2pt	\hbox{\vrule width.#2pt height#1pt \kern#1pt	\vrule width.#2pt}	\hrule height.#2pt}}}}

\documentclass[12pt,a4paper]{article}

\begin{document}
\title{Can rigidly rotating polytropes be sources of the Kerr metric?}
\author{J.\ Mart\'{\i}n${}^1$, A.\ Molina${}^2$ and E.\ Ruiz${}^1$\\
[.5ex]
${}^1$\emph{Departamento de F\'\i sica Fundamental},\\\emph{Facultad de Ciencias, Universidad de Salamanca},\\\emph{Plaza de la Merced s/n, 37008 Salamanca, Spain}.\\
${}^2$\emph{Departament\ de F\'\i sica Fonamental},\\\emph{Universitat de Barcelona},\\\emph{Diagonal 647, Barcelona 08028}.
}
\maketitle\date{}

\begin{abstract}
We use a recent result by Cabezas et al. \cite{Cabezas1} to build up an approximate solution to the gravitational field created by a rigidly
rotating polytrope. We solve the linearized Einstein equations inside and outside the surface of zero pressure including second-order corrections due to
rotational motion to get an asymptotically flat metric in a global harmonic coordinate system. We prove that if the metric and their first derivatives are
continuous on the matching surface up to this order of approximation, the multipole moments of this metric cannot be fitted to those of the Kerr metric.
\end{abstract}

\section{Introduction}

In this article we explore the possibilities arising from the method introduced by Cabezas et al.  \cite{Cabezas1} in a recent paper (hereinafter 
CMMR) for computing approximate solutions of Einstein's equations which can describe the stationary axisymmetric gravitational field of a rigidly rotating
perfect fluid. In their paper, Cabezas et al. study a fluid with a very simple equation of state; the mass-energy density is constant. However, they suggest
that their approximation scheme may also be implemented with more realistic--and more complex--equations of state. Here we use their approach with some minor
changes to deal with a polytropic fluid. However, we do not carry the approximation as far as Cabezas et al. did. The main reason for this is that, in our problem, the
metric cannot be written using elementary functions; it involves a few functions defined by non-trivial differential equations. Therefore, we stop at the level of
the linearized Einstein theory but include quadratic effects in the rotational motion of the fluid. We hope, however, to work out true non-linear terms for this metric in the future.

We aim, first, to show how a polytropic fluid can be incorporated into the CMMR framework and set up the mechanism to run. Secondly, we
use the results the approach provides (which do not differ much from those of Newtonian theory) to question the Kerr metric as a ``good''
metric to describe the gravitational field outside a rigidly rotating polytrope. 

The existence of a suitable source for the Kerr metric that is  a reasonable perfect fluid
  has been discussed at some length. The first result is  due to Roos
\cite{Roos}, who points out that a perfect fluid cannot be ruled out as an 
interior of the Kerr metric by local arguments based on the constraints that matching
imposes on the interior metric: there always exists an interior solution in a neighborhood of the surface of the fluid. However, this argument is insufficient for a
problem which involves elliptic differential equations. Moreover, the proof that Mars and Senovilla \cite{Marc-Seno} and Vera \cite{Raul} give for the uniqueness of the exterior metric surrounding a stationary axisymmetric perfect fluid requires not only the Darmois matching conditions but also an asymptotically flat exterior metric. All of this strengthens the case for the solution to this problem being global, since it implies that there are 
conditions on the metric in domains of the spacetime
manifold which are not close to one another: regularity conditions on the symmetry axis, matching on the fluid surface, and asymptotic flatness at infinity. Even
though we do not actually obtain an exact solution to all these problems, we deal with all of them and solve them in a  coherent way since they are all worked out up to
the order of approximation we are considering. In this sense, the CMMR approach that we follow in this paper casts a reasonable doubt on the existence of a
normal source for the Kerr metric. 

There is a  rather heuristic argument due to Wolf and Neugebauer \cite{Neugebauer} which also suggests that the Kerr metric is not suitable to describe the exterior gravitational field. Although it does not follow the scheme we have sketched above, it too questions perfect fluids as sources of the Kerr metric. For this reason we feel it is worth mentioning.

As we have already mentioned, one of the most important aspects of the CMMR approach is its global character. This arises from several assumptions, one of the
most important of which concerns the coordinates. They are harmonics and cover the entire spacetime manifold in such a way that the metric is of class $C^1$ on the matching
surface (Lichnerowicz matching conditions \cite{Lichnerowicz}). Another important assumption concerns the metric itself. It is assumed that the metric can be expanded in a double
power series of two dimensionless parameters, $\lambda$ and $\Omega$, the first of which takes into account the weakness of the field, and the second is related to the rotation
of the fluid (which is taken to be in rigid motion). This  general framework is implemented by making other complementary assumptions concerning the dependence of the
metric on the spherical coordinates, 
the form of the matching surface, and the expansion of the metric in terms of 
the parameters $\lambda$ and $\Omega$. We aim to provide just enough information about all these assumptions to allow the reader to follow our paper easily. However, for a more detailed
explanation we direct the reader to  the original CMMR article and to a previous paper by Cabezas and Ruiz \cite{Cabezas2}.

In the first three sections of the paper we introduce the polytropic problem guided by  the CMMR approach. Our aim is to
set up the notation and almost everything that is necessary to understand the approximation scheme. We devote Section 4 to solving the interior problem
and Section 5 to matching it to the CMMR exterior solution. The relationship between the global approximate metric we build up and the Kerr metric is discussed in
Section 6. We prove a theorem that excludes the Kerr metric from the set of admissible exterior metrics. In the last section we make some comments and develop an
argument to extend our result to other barotropic equations of state.

\section{Density and pressure}

We consider a Papapetrou-type stationary and axisymmetric metric.
We can therefore choose coordinates $\{t,\,r,\,\theta,\,\varphi\}$ adapted to
the two commuting Killing fields defining the symmetry, and to
the two-dimensional surfaces orthogonal to their orbits. We denote by
$\bm\xi=\partial_t$, the time-like Killing field, and by
$\bm\zeta=\partial_\varphi$, the  azimuthal space-like field, so
the metric can be written as follows:
\begin{eqnarray}
&\bm{g} = \gamma_{tt}\,\bm{\omega}^t{\otimes\,}\bm{\omega}^t
+\gamma_{t\varphi}(\bm{\omega}^t{\otimes\,}\bm{\omega}^\varphi+\bm{\omega}^\varphi{\otimes\,}\bm{\omega}^t)+
\gamma_{\varphi\varphi}\,\bm{\omega}^\varphi{\otimes\,}\bm{\omega}^\varphi
\nonumber\\
&\quad +\,\,\gamma_{rr}\,\bm{\omega}^r{\otimes\,}\bm{\omega}^r+
\gamma_{r\theta}(\bm{\omega}^r{\otimes\,}\bm{\omega}^\theta+\bm{\omega}^\theta{\otimes\,}\bm{\omega}^r) 
+\gamma_{\theta\theta}\,\bm{\omega}^\theta{\otimes\,}\bm{\omega}^\theta\,,
\label{eqmetrica}
\end{eqnarray}
where the $\gamma$'s are functions of $r$ and $\theta$ alone, and
$\bm{\omega}^t=dt$, $\bm{\omega}^r=dr$, $\bm{\omega}^\theta=r\,d\theta$,
$\bm{\omega}^\varphi=r\sin\theta\,d\varphi$.

We want to link this metric to a rigidly rotating perfect fluid. To do so we
consider an energy-momentum tensor,
\begin{equation}
\bm{T} = \left(\mu + p\right)\bm{u}\otimes\bm{u}+p\,\bm{g}\,,
\label{eqenermom}
\end{equation}
invariant under the two Killing fields, so the functions $\mu$ and $p$ (the density and
pressure of the fluid) are also functions of $r$ and $\theta$ alone,
and $\bm{u}$, the velocity of the fluid, is a linear
combination of the two Killing fields, 
\begin{equation}
\bm{u} = \psi\left(\bm\xi + \omega\,\bm\zeta\right)\,,
\label{velocidad}
\end{equation}
where $\omega$ is a constant and $\psi$,
\begin{equation}
\psi \equiv
\left[-\left(\gamma_{tt}+2\omega\,\gamma_{t\varphi}\,r\sin\theta+\omega^2
\,\gamma_{\varphi\varphi}\,r^2\sin^2\theta\right)\right]^{-\frac12}
\label{eqnorma}
\end{equation}
is a normalization factor such that,
$\bm{g}(\bm{u}\,,\bm{u})=-1$.

In the framework defined by
equations (\ref{eqmetrica}), (\ref{eqenermom}) and (\ref{velocidad}),
the energy-momentum conservation law reduces to a couple of first-order linear differential equations \cite{Boyer},
\begin{equation}
\partial_a p - (\mu + p)\partial_a\ln\psi=0
\qquad (a,b,\dots = r\,,\theta)\,.
\label{Euler}
\end{equation}
These can be integrated if the density and pressure of the
fluid are related by a barotropic equation of state, $p=p(\mu)$. For a
polytropic fluid,
\begin{equation}
p=k\mu^{1+\frac1n}\qquad (n>0)\,,
\label{eqestado}
\end{equation}
and a simple calculation leads to the following expressions for $\mu$ and $p$
as functions of the potential $\psi$:
\begin{eqnarray}
&&\mu=\frac1{k^n}\left[\left(\frac\psi\psi_\Sigma\right)^{\frac1{n+1}}-1\right]^{n}\,,
\nonumber\\
[.6ex]
&&p=\frac1{k^n}\left[\left(\frac\psi\psi_\Sigma\right)^{\frac1{n+1}}-1\right]^{n+1}\,.
\label{densidadpresion}
\end{eqnarray}
We have chosen the integration constant $\psi_\Sigma$ in such a way that the pressure
and density of the fluid vanish on the surface
\begin{equation}
\Sigma : \psi(r,\theta)=\psi_\Sigma\,,
\label{implicitsuperficie}
\end{equation}
thus defining the boundary of the fluid.

Equations (\ref{densidadpresion}) and (\ref{implicitsuperficie}) play an important
part in our approach. Both results are exact for a rigidly rotating perfect fluid.

\section{Weak field approximation}

We look for a solution to the problem set up in the preceding section. We assume it
takes the form of a metric $\bm{g}(\lambda,\Omega)$ depending on two
dimensionless parameters
$\lambda$ and $\Omega$ having the following properties: 
\begin{enumerate}
\item $\bm{g}(\lambda,\Omega)$ tends to the Minkowski
metric if $\lambda$ goes to $0$, that is,
\begin{equation}
\bm{g}(0,\Omega)=\bm{\eta}=-\bm{\omega}^t{\otimes\,}\bm{\omega}^t
+\bm{\omega}^r{\otimes\,}\bm{\omega}^r+\bm{\omega}^\theta{\otimes\,}\bm{\omega}^\theta
+\bm{\omega}^\varphi{\otimes\,}\bm{\omega}^\varphi\,;
\label{Minkowski}
\end{equation}
we use this limit to identify coordinates $\{t,r,\theta,\varphi\}$ as the
standard polar coordinates of flat spacetime; 
\item $\bm{g}(\lambda,\Omega)$ tends to a
static spherically symmetric metric when $\Omega$ goes to $0$, which is 
a solution of
Einstein's equations for a fluid with the same equation of state, say a polytrope, as
that of the stationary axisymmetric solution.
\end{enumerate}

The parameter $\Omega$ must be related to the rotational motion of
the fluid. Following CMMR, we assume 
$\omega=\lambda^{1/2}\Omega r_0^{-1}$, where $r_0$ is a constant with the dimensions of length  that may be
identified as the radius of the fluid ball in the static limit. The other parameter
$\lambda$ obviously  accounts for 
the gravitational field strength. Though we cannot
yet identify it with any definite quantity (but we suppose that it may be proportional to
the quotient of the ``mass'' of the fluid and its ``radius'') we are sure about the
role it plays in the approximation scheme: we expect the metric to
behave as follows for small values of $\lambda$:
\begin{eqnarray}
&&\gamma_{tt} \approx -1 + \lambda f_{tt}\,,\quad
\gamma_{t\varphi} \approx\lambda^{3/2}\Omega f_{t\varphi}\,,\quad
\gamma_{\varphi\varphi} \approx 1+\lambda f_{\varphi\varphi}\,,
\nonumber\\
&&\gamma_{rr} \approx 1 + \lambda f_{rr}\,,\quad 
\gamma_{r\theta} \approx\lambda f_{r\theta}\,,\quad
\gamma_{\theta\theta} \approx 1+\lambda f_{\theta\theta}\,.
\label{aproxlineal}
\end{eqnarray}
These expressions agree with the two conditions we impose on the metric and they
also give a more precise meaning to the kind of approximation we are proposing.

The above expansion for the metric in $\lambda$ leads to a similar expansion of the
energy-momentum tensor. Taking into account (\ref{aproxlineal}), it can easily be checked
that all the quantities entering in
$\bm{T}$ except $\mu$ and $p$ have non-zero values at $\lambda=0$:
$\bm{g}\approx\bm{\eta}$, $\psi\approx 1$ and $\bm{u}\approx -\bm{\omega}^t$ (we use the
same symbol to denote the vector field and the $1$-form). However, a coherent
perturbation scheme based on the parameter $\lambda$ needs an energy-momentum tensor
which tends to zero with the first power of $\lambda$. This can only be
achieved if the density and the pressure have a linear term in $\lambda$. This does not seem evident.

The normalization factor up to first order in $\lambda$ reads:
\begin{equation}
\psi \approx 1 + \frac{1}{2}\lambda \left(f_{tt} +\Omega^2
\eta^2\sin^2\theta\right)\,,
\label{psiapprox}
\end{equation}
where $\eta=r/r_0$. Since the constant $\psi_\Sigma$ is equal to the value of
$\psi$ on the zero pressure surface, we can assume a similar expansion for it in 
$\lambda$ and
$\Omega$, so we write
\begin{equation}
\psi_\Sigma\approx 1+\lambda\left[M_0
+\Omega^2\left(\frac{1}{3}-M_0\kappa\right)\right]\,.
\label{psisigma}
\end{equation}
This may seem rather bizarre, but it is just a way to introduce two constants $M_0$ and
$\kappa$ (the reason for this choice will be given later). Substituting
expressions (\ref{psiapprox}) and (\ref{psisigma}) into equation (\ref{densidadpresion})
and expanding in $\lambda$, we get
\begin{eqnarray}
&&\mu\approx\frac{\lambda^n}{k^n(n+1)^n}
\left[\frac{1}{2} \left(f_{tt}-2M_0\right) +\Omega^2\left(\frac{1}{2}\eta^2\sin^2\theta
-\frac{1}{3}+M_0\kappa\right)\right]^n\nonumber\\
&&\quad \equiv\frac{\lambda^n}{k^n(n+1)^n}\,q^n\,,\label{densidadaprox}\\
[.6ex]
&& p\approx \frac{\lambda^{n+1}}{k^n(n+1)^{n+1}}\,q^{n+1}\,.
\label{presionaprox}
\end{eqnarray}
This shows that we can make $\mu$ a quantity of first order in
$\lambda$ if we assume $k\propto\lambda^{1-\frac{1}{n}}$; a suitable choice
that takes account of physical density units and other technical facts related to the matching is
\begin{equation}
\frac{1}{k^n}=\frac{(n+1)^n}{4\pi r_s}\lambda^{1-n}\,,
\label{deflambda}
\end{equation}
where $r_s$ is a new length constant. This equation can actually be
a definition of the parameter $\lambda$ whenever we can establish a relationship between the
length scales, $r_0$ and $r_s$. We shall do this in Section \ref{secenganche}. 

We now come to the field equations. We do not aim to work out an exact solution but rather an
approximate one. Following CMMR once more, we use the so-called post-Minkowskian
approximation scheme (see for instance \cite{MTW} for a general description of the
approach). It requires the introduction of 
 new coordinates
$\{t,\,x=r\sin\theta\cos\varphi,\,y=r\sin\theta\sin\varphi,\,
z=\cos\theta\}$; standard Cartesian coordinates associated with the spherical-type
coordinates $\{t,\,r,\,\theta,\,\varphi\}$, and the quantity
\begin{equation}
h_{\alpha\beta}\equiv  g_{\alpha\beta}-\eta_{\alpha\beta}\,,
\label{desviavion}
\end{equation}
which is assumed to be of at least first order in $\lambda$. Here the indexes
$\alpha,\beta\dots$ stand for the new coordinates, and expression (\ref{desviavion}) relates
the components of the spacetime metric and the Minkowski metric in this system
of coordinates. That is, $(\eta_{\alpha\beta})={\rm diag}(-1,1,1,1)$ and
$g_{\alpha\beta}$ are some combinations of the $\gamma$'s introduced in (\ref{eqmetrica}). These assumptions are
consistent with those made before in terms of the old coordinates (equation (\ref{aproxlineal})). Moreover, we require the new Cartesian-like
coordinates to be harmonic coordinates.

Under the assumptions we have been making concerning the metric and the coordinates up to now,
the linearized Einstein equations imply that
$h_{\alpha\beta}$ must be a solution of the differential system
\begin{eqnarray}
&&\triangle h_{00} = -2\frac{\lambda}{r_s^2}q^n\,,\label{eqint00}\\
[.6ex]
&&\triangle h_{0i} = 4\frac{\lambda^{3/2}\Omega}{r_s^2}q^n\eta \sin\theta\,m_i\,,
\label{eqint0i}\\
[.6ex]
&&\triangle h_{ij} = -2\frac{\lambda}{r_s^2}q^n\delta_{ij}\,,\label{eqintij}\\
[.6ex]
&&\partial^k (h_{k\mu} -\frac12
h\,\eta_{k\mu}\,) = 0\,,
\label{eqintharmo}
\end{eqnarray}
where $\triangle\equiv\delta^{ij}\partial_i\partial_j$ stands for the standard Laplacian
in Cartesian coordinates, $h\equiv\eta^{\gamma\mu}h_{\gamma\mu}$, and $(m_i)\equiv
(-\sin\varphi,\cos\varphi,0)$. 
As defined in
(\ref{densidadaprox}), $q$  depends on $f_{tt}$, which is not a great problem since
$h_{00}\approx \lambda f_{tt}$. Interestingly the
pressure does not contribute to the right-hand side of equations
(\ref{eqint00}), (\ref{eqint0i}), or (\ref{eqintij}). It can easily be checked from 
definition (\ref{deflambda}) and formula (\ref{presionaprox}) that $p$ is a second-order quantity in $\lambda$.

\section{Slow rotation solution}

In this section we work out an approximate metric which describes the
geometry of the spacetime we are interested in, which 
involves the rotation parameter $\Omega$. We do not intend to find an exact solution
to the linear problem defined by equations (\ref{eqint00}) to (\ref{eqintharmo}), but
rather an approximate solution up to order $\Omega^2$. The differential equations used to describe this system include
one that should verify the component $h_{00}$, which seems to be the most difficult to
solve, since $h_{00}$ appears on the right-hand side 
of the equation in a non-trivial way. So, first we deal with this equation and
later we try to find a solution for the whole system.

Let us set $\Omega=0$ in (\ref{densidadaprox}). Substituting the result of
(\ref{eqint00}) and replacing
$h_{00}$ by $\lambda f_{tt}$ (so we eliminate any dependence on $\lambda$) gives the following equation for the time-time component:
\begin{equation}
\triangle f_{tt}=-\frac{2}{r_s^2}\left(\frac{1}{2}
f_{tt}-M_0\right)^n\,.
\end{equation}
We can assume spherical symmetry in this limit; that is, $f_{tt}(r)$. Then, by introducing
a new function,
\begin{equation}
f_{tt}(r)=2\left[\Phi\left(\frac{r}{r_s}\right)+M_0\right]\,,
\label{h00cero}
\end{equation}
and changing the independent variable, $r=r_s s$, we arrive at the Lane-Emden equation of Newtonian theory \cite{Chandrasekhar} (see also \cite{Weinberg} and \cite{Hansen}),
\begin{equation}
\Phi''+\frac{2}{s}\Phi'+\Phi^n=0\qquad \left('\equiv
\frac{d}{ds}\right)\,.
\label{LE}
\end{equation}
We need a solution of this equation which is regular at the origin of the coordinates, $s=0$. We can select it by looking at well-known results of classical polytrope theory: the solution of the Lane-Emden equation that satisfies the initial data $\Phi(0)=1$ and $\Phi'(0)=0$ is chosen (see \cite{Chandrasekhar}).
Hereafter we shall identify the symbol $\Phi$ with that particular solution of the Lane-Emden equation. It is important to remember that there are only analytic expressions for
$\Phi$ for a few  polytropic indexes, $n=0$ (constant mass density), $n=1$ (the Lane-Emden equation is linear), and $n=5$
\cite{Chandrasekhar}. In all other cases $\Phi$ has to be calculated using numerical integration methods.

We now have an approximation of $h_{00}$ which is of zeroth order in $\Omega$. 
 To obtain a second-order correction, we replace $h_{00}$ by
\begin{equation}
h_{00}(r,\theta)\approx \lambda f_{tt}(r,\theta)\ \rightarrow\
2\lambda\left[\Phi(s)+M_0\right]+\lambda\Omega^2
f_{tt}(r,\theta)\,.
\label{h00dosprop}
\end{equation}
Substituting this expression into (\ref{eqint00}), 
expanding this equation to include all the quadratic terms in $\Omega$, and 
taking into account that $\Phi$ verifies (\ref{LE}), we get the following linear
equation for the new $f_{tt}$:
\begin{equation}
\triangle f_{tt}+\frac{n}{r_s^2}\Phi^{n-1}f_{tt}=
-\frac{2n}{r_s^2}\Phi^{n-1}\left(\frac12\eta^2\sin^2\theta-\frac13+M_0\kappa\right)\,.
\label{f2solucion}
\end{equation}
This does not appear easy to solve. However, since $\Phi$ does not depend on the
$\theta$ coordinate, we can look for a solution of the homogeneous equation associated with
(\ref{f2solucion}) as a function of $r$ times a Legendre polynomial. The
CMMR approach \cite{Cabezas1} sets the rule that a term of
order $\Omega^2$ in the
$g_{00}$ component of the metric should not depend on a Legendre polynomial of
higher than second order. If we accept this rule, the following function may be
enough for our purposes:
\begin{equation}
f_{tt}(r,\theta)=\frac{4r_s^2}{r_0^2}\phi_0(s)
+\frac23(1-\eta^2)-2M_0\kappa
+\left[\phi_2(s)+\frac23\eta^2\right]P_2(\cos\theta)\,,
\label{h00dos}
\end{equation}
where $\phi_0(s)$ and $\phi_2(s)$ are, respectively, solutions of the following
two linear differential equations,
\begin{eqnarray}
&&\phi_0''+\frac2s\phi_0'+n\Phi^{n-1}\phi_0=1\,,\nonumber \\
[.6ex]
&&\phi_2''+\frac2s\phi_2'+\left(n\Phi^{n-1}-\frac6{s^2}\right)\phi_2=0\,.
\label{difecuacion}
\end{eqnarray}
Both equations admit a one-parameter family of solutions which are
regular at $s=0$. We refer to
these regular solution when we write the symbols $\phi_0$ and $\phi_2$.

We now use the approximate solution for $h_{00}$ given by (\ref{h00cero}), (\ref{h00dosprop}),  and(\ref{h00dos}) to expand the right-hand side of 
(\ref{eqint0i}). Eliminating 
 terms in $\Omega^3$ amounts to substituting $q$ by $\Phi$, and gives us:
\begin{equation}
\triangle h_{0i}=4\frac{\lambda^{3/2}\Omega}{r_s^2}\Phi^n\eta \sin\theta\,m_i\,.
\end{equation}
According to the CMMR prescription, the solution of this equation that we need has two parts: a specific solution of the
full equation and a solution  of the homogeneous equation that depends on the angular coordinates through a 
spherical harmonic vector, that is: 
\begin{equation}
h_{0i}(r,\theta,\varphi)=- \lambda^{3/2}\Omega\left[j_1+4\left(\Phi(s)-\frac13-\frac{2r_s^2}{r_0^3\eta^3}I(s)\right)\right]\eta\sin\theta\, m_i\,,
\label{h0iuno}
\end{equation}
where $j_1$ is a constant and
\begin{equation}
I(s)\equiv \int_0^s \tau^2\Phi(\tau)d\tau\,.
\end{equation}
This solution is regular at $s=0$ and it also satisfies the harmonic condition (\ref{eqintharmo}).

The equations involving the components $h_{ij}$ are easier to solve. Comparing (\ref{eqintij}) with (\ref{eqint00}), it is obvious that $h_{00}\,\delta_{ij}$ is a
specific solution of the first equation if $h_{00}$ is a solution of the second one. Moreover, the couple $h_{00}$ and $h_{ij} $ so defined always satisfies the
harmonic condition  (\ref{eqintharmo}). We may add to $h_{ij}$ any solution of the homogeneous equation and the harmonic condition; that is, 
the terms associated with the constants $a_0$, $b_0$, $a_2$ and $b_2$, but not $m_2$, in equation (33) of CMMR. However, since they are not necessary to match this solution to the exterior one, we omit them.

Finally, bringing together the results obtained in this section and joining the components of the approximate interior metric in a tensor
expression we have:
\begin{eqnarray}
&&\bm{g}_{\rm int} \approx
\left(-1+2\lambda\Phi+2\lambda M_0\right) \bm{T}_0
+\left(1+2\lambda\Phi+2\lambda M_0\right) \bm{D}_0\nonumber\\
&&\qquad\quad + \lambda\Omega^2\left[\frac{4r_s^2}{r_0^2}\phi_0+\frac23(1-\eta^2)-2M_0\kappa\right]
(\bm{T}_0+\bm{D}_0)\nonumber\\
&&\qquad\quad + \lambda\Omega^2\left(\phi_2+\frac23\eta^2\right)(\bm{T}_2+\bm{D}_2)\nonumber\\
&&\qquad\quad + \lambda^{3/2}\Omega\eta\left[j_1+4\left(\Phi-\frac13-\frac{2r_s^2}{r_0^3\eta^3}I\right)\right]\bm{Z}_1\,,
\label{gint}
\end{eqnarray}
where we have introduced the CMMR notations to denote spherical harmonic
tensors, $\bm{T}_0\equiv\bm{\omega}^t\otimes\,\bm{\omega}^t$, $\bm{D}_0\equiv \delta_{ij}dx^i{\otimes\,}dx^j$, $\bm{T}_2\equiv P_2(\cos\theta)\,\bm{\omega}^t\otimes\,\bm{\omega}^t$, $\bm{D}_2\equiv P_2(\cos\theta)\,\delta_{ij}dx^i{\otimes\,}dx^j$
and $\bm{Z}_1\equiv P_1^1(\cos\theta)\,(\bm{\omega}^t\otimes\bm{\omega}^\varphi
+\bm{\omega}^\varphi\otimes\bm{\omega}^t)$.

\section{Global metric\label{secenganche}}

The next step is to connect the interior metric (\ref{gint}), through the zero pressure surface defined by (\ref{implicitsuperficie}), with an asymptotically flat
vacuum metric. To accomplish this we use the CMMR exterior metric,
\begin{eqnarray}
&&\bm{g}_{\rm ext} \approx
\left(-1+2\lambda\frac{M_0}{\eta}\right)\bm{T}_0
+\left(1+2\lambda\frac{M_0}{\eta}\right)\bm{D}_0 \nonumber\\
&&\qquad\quad
+2\lambda\Omega^2\frac{M_2}{\eta^3}\left(\bm{T}_2+\bm{D}_2\right)
+2\lambda^{3/2}\Omega\frac{J_1}{\eta^2}\,\bm{Z}_1\,,
\label{gext}
\end{eqnarray}
where we have set the
constants appearing in the original CMMR expression equal to zero. This does not
imply any lack of generality in our approach, since including or omitting these terms is just a matter of convenience for matching correctly the interior and exterior metrics.

Since we have an approximate expression for the interior metric, we can use it to get a parametric equation for the matching surface (\ref{implicitsuperficie}).
However, it is preferable to do this using the exterior metric, since we assume that the metric is
continuous over the zero pressure surface, as in CMMR. This leads to
\begin{equation}
\Sigma : r \approx r_0\left[1+ \Omega^2\kappa+\frac{\Omega^2}{M_0}\left(M_2-\frac13\right)P_2(\cos\theta)\right]\,,
\label{parasuperficie}
\end{equation}
and also to the expression for $\psi_\Sigma$ which we have been using (\ref{psisigma}). We have included a new constant, $\kappa$, which was absent in the CMMR
expression for the parametric equation of the matching surface. Even though it was superfluous there, we need it to solve the matching problem
here.

We understand the matching of interior and exterior metrics in the same way as Lichnerowicz does: the metric and its derivatives must be continuous on the matching
surface \cite{Lichnerowicz}.  However, since we do not have an exact solution, we require these conditions to be fulfilled up to the same order of approximation as the
metric is a solution of the field equations. This means that we evaluate both metrics on the surface defined by (\ref{parasuperficie}), then we expand the result
in $\Omega$ neglecting terms of a higher order than $\Omega^2$, then we develop the matching conditions in spherical harmonic tensors, and finally we equate all the
coefficients of the expansion to zero to get a set of algebraic equations. This procedure leads to the
following six constraints,
\begin{eqnarray}
&& \Phi\left(\frac{r_0}{r_s}\right)+\Omega^2\left[\frac{2r_s^2}{r_0^2}\phi_0\left(\frac{r_0}{r_s}\right)
+\frac{r_0}{r_s}\kappa\Phi'\left(\frac{r_0}{r_s}\right)\right]\approx 0\,,\nonumber\\
[.6ex]
&& M_0+\frac{r_0}{r_s}\Phi' \left(\frac{r_0}{r_s}\right)
+\Omega^2 \left[\frac{2r_s}{r_0} \phi_0' \left(\frac{r_0}{r_s}\right)+\frac{r_0^2}{r_s^2}\kappa\Phi'' \left(\frac{r_0}{r_s}\right)-\frac23-2M_0\kappa \right]\approx 0\,,\nonumber\\
[.6ex]
&& \phi_2\left(\frac{r_0}{r_s}\right)+\frac{2r_0}{r_sM_0}\left(M_2-\frac13\right)\Phi' \left(\frac{r_0}{r_s}\right)\approx 0\,,\nonumber\\
[.6ex]
&& M_2+\frac43+\frac{r_0}{2r_s}\phi_2' \left(\frac{r_0}{r_s}\right)+
\frac{r_0^2}{r_s^2M_0}\left(M_2-\frac13\right)\Phi'' \left(\frac{r_0}{r_s}\right)\approx 0\,,\nonumber\\
[.6ex]
&& j_1-2J_1-\frac43-\frac{8r_s^3}{r_0^3}I\left(\frac{r_0}{r_s}\right)+
4 \Phi\left(\frac{r_0}{r_s}\right)\approx 0\,,\nonumber\\
[.6ex]
&& j_1+4J_1-\frac43+\frac{16r_s^3}{r_0^3}I\left(\frac{r_0}{r_s}\right)-
4 \Phi\left(\frac{r_0}{r_s}\right)+4\frac{r_0}{r_s} \Phi' \left(\frac{r_0}{r_s}\right)\approx 0\,.
\label{enganche}
\end{eqnarray}
One may expect a large number of constraints, but as $h_{ij}$ is essentially equal to $h_{00}$, all the constraints we can derive from the matching of these
components are already included in the matching of $h_{00}$.

There are two types 
of constant in the system (\ref{enganche}). Two which are of the first type, $r_0/r_s$ and $M_0$, come from the static limit; the others, of the second type, appear at a level
where rotation is taken into account. This last class can be taken as pure numbers when we try to solve the matching conditions but this is not the case for the first class, which may be linear functions of $\Omega^2$ (constants are seen as functions of $\lambda$ and $\Omega$ in the CMMR approach). However,  in our
problem we mean that it  is better to consider $r_s$ as a true constant and to assume that there is no $\Omega^2$ term in the expansion of $M_0$. This leads to
\begin{equation}
\Phi\left(\frac{r_0}{r_s}\right)= 0\,,\qquad
M_0\approx -\frac{r_0}{r_s}\Phi' \left(\frac{r_0}{r_s}\right)\,;
\label{Newcondicion}
\end{equation}
two well-known predictions of Newtonian theory \cite{Chandrasekhar}. In order to justify our choice: first, recall expression
(\ref{deflambda}), which becomes an interesting definition of the parameter $\lambda$ if $r_s$ is not a free constant.
The first equation in (\ref{Newcondicion}) ensures this since it permits $r_0/r_s$ to be determined in terms of the zeroes of $\Phi(s)$. So we can write,
$r_s=r_0/s_0$, where $s_0$ is a number
\footnote{A classical theorem proves that the solution of the Lane-Emden equation corresponding to the initial data $\Phi(0)=1$ and $\Phi(0)=0$ has a first zero in
the interval
$0<s<\infty$ if
$0\leq n<5$ \cite{Chandrasekhar}.}. Another technical reason is based on a careful inspection of the first matching  constraint in
(\ref{enganche}). It seems that any dependence on $\Omega$ we assign to $r_s$ may be absorbed into the  extra constant $\kappa$. Second, we do not expect
any correction to the mass at the Newtonian level; this is 
the meaning of our assumption concerning $M_0$, the monopole moment of the exterior gravitational field.

Taking into account (\ref{Newcondicion}), we can solve the matching conditions (\ref{enganche}) to get approximate expressions for the first multipole moments of the metric,
\begin{equation}
M_2 \approx \frac13+\frac12\phi_2(s_0)\,,\qquad
J_1\approx \frac23M_0-\frac4{s_0^3}I(s_0)\,,
\label{momentos}
\end{equation}
the constants of the interior metric,
\begin{equation}
\kappa\approx 2\frac{\phi_0(s_0)}{s_0^2M_0}\,, \qquad
j_1\approx \frac43(1+M_0)\,,
\label{intconstantes}
\end{equation}
and the initial data needed to pick out
regular solutions of the differential equations (\ref{difecuacion}),
\begin{equation}
 \phi'_0(s_0)\approx \frac{s_0}{3}\,, \qquad
 \frac12s_0\phi'_2(s_0)+\frac32\phi_2(s_0)+\frac53 \approx 0\,.
 \label{inicial}
\end{equation}

Substituting (\ref{momentos}) and (\ref{intconstantes}) into expressions (\ref{gint}) and (\ref{gext}), we get an approximate solution of Einstein's equations
inside and outside the polytropic fluid up to order $\lambda^{3/2}$ and $\Omega^2$, which is of class $C^1$ on the surface of the fluid
up to the same order of approximation if functions $\phi_0(s)$ and $\phi_2(s)$ satisfy the conditions (\ref{inicial}).

\section{Kerr metric}

The vacuum metric (\ref{gext}) can be used to describe the Kerr metric near infinity by choosing suitable values for the constants $M_0$, $J_1$, and $M_2$. This is a
straightforward consequence of multipole moment theory in harmonic coordinates \cite{Thorne}. So, to the extent that a stationary axisymmetric vacuum metric can be
identified by its multipole moments \cite{Walter}, \cite{Kundu} we can say that the metric (\ref{gext}) coincides with the Kerr metric if: \footnote{For a development of Kerr multipole moments that is
closer to our point of view see 
\cite{ABMMR}}
\begin{equation}
m=\lambda  r_0M_0\,, \qquad ma=\lambda^{3/2}\Omega r_0^2J_1\,, \qquad
 -ma^2=\lambda\Omega^2r_0^3M_2\,,
 \label{Kerr}
 \end{equation}
where $m$ and $a$ are the standard parameters of the Kerr metric in Boyer-Lindquist coordinates. However, if we consider the metric  (\ref{gext}) to be the exterior
gravitational field of a polytrope, the matching constraints restrict the values that $M_0$, $J_1$ and $M_2$ can take. Therefore, we may ask if we still have
enough freedom to make the exterior metric into the Kerr metric.

Solving the  first two expressions in (\ref{Kerr}) for
$m$ and $a$, and substituting the results 
into the third one, we find that the Kerr quadrupole
moment,
$-ma^2$, is a quantity of order $\lambda^2$ not of order $\lambda$ as it appears on right-hand side of the equation. This condition cannot be fulfilled unless
$M_2$ is a quantity of order $\lambda$, $M_2\approx 0$. 
Equations (\ref{momentos}) and
(\ref{inicial}) transform this condition on $M_2$ into two initial data for the function $\phi_2(s)$
\begin{equation}
\phi_2(s_0)\approx  -\frac23\,, \qquad \phi_2'(s_0)\approx -\frac{4}{3s_0}\,.
\label{Kerrcondi}
\end{equation}
We must add to them the regularity condition $s=0$. This means there are too many conditions for a function defined by a linear second-order differential equation. Let us prove
that it is not possible.
\medskip

 {\bf Theorem.} {\it Let $\Phi(s)$ be the solution of the Lane-Emden equation defined by the initial data $\Phi(0)=1$, $\Phi'(0)=0$, and let $s_0$ be the first
zero of $\Phi(s)$ in the interval $(0,+\infty)$, that is  $\Phi(s_0)=0$. The differential equation
 \begin{equation}
 \left(s^2\phi_2'\right)'+\left(ns^2\Phi^{n-1}-6\right)\phi_2=0\,,
 \label{difecu}
 \end{equation}
 has no smooth solution in the interval $(0, s_0)$ such that
  \begin{equation}
 \phi_2(0)=0\,, \qquad \phi_2(s_0)=-\frac{2}{3}\,,\qquad \phi_2'(s_0)=-\frac{4}{3s_0}\,,
 \label{iniciales}
 \end{equation}
 }

To say that $\phi_2(s)$ vanishes at $s=0$ is equivalent to saying that $\phi_2(s)$ is a regular solution of equation (\ref{difecu}). This implies $\phi_2(s)\sim
c_2s^2$ ($c_2\neq 0$) in a neighborhood 
of the origin. Then, to prove the theorem we have to demonstrate that any solution of this kind can take the
values (\ref{iniciales}) at $s_0$. Let us first prove the following:
\medskip

{\bf Lemma.} {\it If a solution $\phi_2(s)\neq 0$ of the differential equation (\ref{difecu}) is regular at $s=0$, then it does not vanish in the interval
$(0,s_0)$.}
\smallskip

Let us consider the differential equation,
\begin{equation}
 \left(s^2\phi_1'\right)'+\left(ns^2\Phi^{n-1}-2\right)\phi_1=0\,.
\label{dipolo}
\end{equation}
Its regular solutions at $s=0$ can be written in terms of the Lane-Emden solution, $\phi_1(s)=c_1\,\Phi'(s)$ ($c_1\neq 0$). If $c_1>0$, $\phi_1(s)$ is negative in
$(0,s_0)$ because $\Phi(s)$ is a decreasing function of $s$ in that domain. Let us set $c_1=1$.

Multiplying (\ref{difecu}) by $\Phi'(s)$ and (\ref{dipolo}) by $\phi_2(s)$, then subtracting them, and integrating the result over $(0,s)$, we get
\begin{equation}
W(s)\equiv \Phi'(s)\phi_2'(s)-\Phi''(s)\phi_2(s)=\frac{4}{s^2}\int_0^s\Phi'(\tau)\phi_2(\tau)d\tau\,.
\label{lema}
\end{equation}
The integral on the right-hand side is negative near $s=0$ if $c_2>0$ and positive if $c_2<0$. Let us take $c_2>0$ (the argument runs the same for
$c_2<0$). Since $\Phi'(s)$ is negative,  the product under the integral symbol is negative in $(0,s_0)$ unless $\phi_2(s)$ vanishes. Let us assume
$\phi_2(s_*)=0$, $0<s_*<s_0$. Then, since $\phi_2'(s_*)\leq 0$, we have $W(s_*)\geq 0$, but the right-hand side of (\ref{lema}) is still negative at $s_*$!
Therefore, we conclude that all the zeroes of any regular $\phi_2(s)$ should be outside the interval $(0,s_0)$.
\smallskip

Let us return to the proof of the theorem. We introduce a new function,
\begin{equation}
y(s)\equiv s^6\left[\frac{\phi_2'(s)}{\phi_2(s)}-\frac{2}{s}\right]\,.
\label{defy}
\end{equation}
If $\phi_2(s)$ is a solution of (\ref{difecu}), $y(s)$ is a solution of the first-order  differential equation
\begin{equation}
y'+\frac{y^2}{s^6}+ns^6\Phi^{n-1}=0\,.
\label{nueva}
\end{equation}
Clearly $y(s)$ is the same function for all regular solutions of (\ref{difecu}), particularly 
the solution of (\ref{nueva}) defined by the initial data
$y(0)=0$. The other two conditions in  (\ref{iniciales}) lead to a further one on $y(s)$: $y(s_0)=0$. The second and third terms in equation (\ref{nueva}) are
positive in $(0,s_0)$, so $y(s)$ is a decreasing function there. Also $y(s)$ 
is negative near $s=0$ if $y(0)=0$, and it is negative in $(0,s_0)$ if it does not
run away to minus infinity for an intermediate value of $s$. From the definition of $y(s)$ and the preceding Lemma, we know that this does not happen.
Therefore $y(s)<0$ at $s_0$ so 
$y(s_0) \neq 0$ and (42) is not fulfilled.

\section{Comments}

We have given an proof based on a perturbation approach which rejects the Kerr metric as the exterior metric of a polytropic fluid. Even though we do not go further
than the linear approximation, but consider  quadratic terms in the rotation,  the approximation surprisingly seems to be enough to come to a conclusion by
analyzing the constraints that the Kerr metric imposes on the quadrupole moment of the exterior field. To obtain our result we need to deal with a global metric
written in a global coordinate system, so we understand the matching between the interior metric and the exterior metric in the sense that the metric components and
their first derivatives must be
continuous on the matching surface. On this basis one can argue against the generality of our result. There are still important questions that we cannot answer: Is it possible to get a different
global metric by using a less restrictive coordinate condition?. Would such a metric admit the Kerr metric as a suitable exterior metric?. Our global metric does, however, contain the desired number of free constants:
$\omega$, which takes account of the rotation; and $r_0$, which may be related to the central density by means of equations (\ref{deflambda}),
\begin{equation}
\mu(0)\approx \frac{\lambda}{4\pi r_0^2}\left[s_0^2+2n\Omega^2\phi_0(0)\right]\,,
\end{equation}
since constraint (\ref{inicial}) implies that $\phi_0(0)$ is a function of $s_0$. It is widely thought that such a two-parameter metric is the general solution to
this problem in Einstein's theory as well as in Newtonian gravity.

Let us give an argument which may extend our results to perfect fluids with other barotropic equations of state which are not of the polytropic kind. The
integrability condition of Boyer's equations (\ref{Euler}) can be fulfilled by setting the density $\mu$ to be a function of the normalization factor $\psi$.
Introducing the new variable $X=\psi/\psi_\Sigma$, we can write $\mu=f(X)$, and the pressure is given by
\begin{equation}
p(X)= X \int_1^X\chi^{-2}f(\chi)d\chi\,.
\label{otrapresion}
\end{equation}
Then $X=1$ defines the fluid boundary, $p(1)=0$. It is clear that $p$ is a function of $\mu$, at least in a local sense, so the fluid admits a barotropic equation
of  state. We have seen (equations (\ref{psiapprox}) and (\ref{psisigma})) that to expand the metric in $\lambda$ implies $X\approx 1+\lambda q$. Therefore, what is
relevant to our approach is the form that the function $f(X)$ takes near $X=1$. For instance, let $f(X)=\mu_sg(X)(X-1)^n$ ($n\geq 0$), where
$g(X)$ is a well-behaved function at $X=1$, $g(1)=1$, and $\mu_s$ is a constant we introduce for convenience to make $g(X)$ dimensionless. Then we have $\mu\approx
\mu_s(X-1)^n\left[1+o(X-1)\right]$, for small values of $X-1$, and $\mu \approx \mu_s\lambda^nq^n$, for the first 
 term in the $\lambda$ expansion of $\mu$. This
expression is similar to the one we obtain for the density of a polytropic fluid (\ref{densidadaprox}), though it will lead to a different value of the parameter
$\lambda$. Furthermore, we can integrate (\ref{otrapresion}) by using the form of $f(X)$ for small values of $X-1$, and by expanding 
the result in
$\lambda$ to check that the pressure is at least one order higher in $\lambda$ than the density. All of this suggests the existence of a common metric up to
first order in $\lambda$ for all these barotropic fluids.

Finally, let us note that if the density is not zero at $X=1$ ($n=0$), the resultant first-order metric must coincide with the CMMR metric. It was shown in the
CMMR article that the Kerr metric does not fit the exterior metric of a constant density
fluid, and neither does it fit the exterior field of this other kind of barotropic fluid.

\section*{Acknowledgments}

This work has been partially supported by projects FIS2006-05319 and FIS2007-63034 from 
the \textit{Ministerio de Educaci\'on y Ciencia} and by project SA010C05
from the \textit{Junta de Castilla y Le\'on}.

\end{document}